# In-situ synthesis and characterization of polyaniline -$CaCu_3Ti_4O_{12}$ nano crystal composites

P.Thomas[a,b], K.Dwarakanath,[a] and K.B.R.Varma.[b*]

[a] Dielectric Materials Division, Central Power Research Institute, Bangalore: 560 080, India.

[b] Materials Research Centre, Indian Institute of Science, Bangalore: 560012, India,

Abstract

High dielectric constant (ca.$2.4 \times 10^6$ @1kHz) nano composite of Polyaniline (PANI) / $CaCu_3Ti_4O_{12}$ (CCTO) was synthesized using a simple procedure involving in-situ polymerization of aniline in dil. HCl. The PANI and the composite were subjected to X-ray diffraction, Fourier transform infrared, Thermo gravimetric, Scanning electron microscopy and Transmission electron microscopy analyses. The presence of the nano crystallites of CCTO embedded in the nanofibers of PANI matrix was established by TEM. Frequency dependent characteristics of the dielectric constant, dielectric loss and AC conductivity were studied for the PANI and the composites. The dielectric constant increased as the CCTO content increased in PANI but decreased with increasing frequency (100Hz – 1MHz) of measurement. The dielectric loss was two times less than the value obtained for pure PANI around 100Hz. The AC conductivity increased slightly upto 2kHz as the CCTO content increased in the PANI which was attributed to the polarization of the charge carriers.

## 1. Introduction

---

[*] Corresponding author : Tel. +91-80-2293-2914; Fax: +91-80-2360-0683.
E-mail : kbrvarma@mrc.iisc.ernet.in (K.B.R.Varma)

Polymeric materials associated with high dielectric constant, high breakdown strength and low dielectric loss have been in increasing demand for the high energy density and pulsed energy capacitors for both military and civilian applications. [1-3]. In order to enhance the dielectric constant in polymers, ceramics such as Pb(Mg$_{1/3}$Nb$_{2/3}$)O$_3$-PbTiO$_3$(PMNT), Pb(Zr,Ti)O$_3$(PZT), BaTiO$_3$ (BT) [4-9] were used as fillers due to their high dielectric constant. The CaCu$_3$Ti$_4$O$_{12}$ (CCTO) ceramic which has centrosymmetric *bcc* structure (space group Im3, lattice parameter a ≈ 7.391 Å, and Z=2, has gained considerable attention due to its unusual high dielectric constant (ε ~ 10$^{4-5}$) which is nearly independent of frequency (upto 10 MHz) and low thermal coefficient of permittivity (TCK) over a wider range of temperature between 100-600K [10,11]. Several schools of thoughts exist to explain the origin of high dielectric constant observed in CCTO ceramics [10-17]. Though several explanations have been put forward, the actual mechanism of the origin of giant dielectric constants in CCTO is still debatable as to whether it is intrinsic or extrinsic in nature. Recently, CaCu$_3$Ti$_4$O$_{12}$ (CCTO) ceramic has been used as a filler and studied to explore the possibility of obtaining high dielectric constant composites for potential capacitor applications [18-22]. It was reported that, the dielectric constant as high as 740 at 1kHz was achieved for a composition of fixed concentration : 50 vol % CCTO and 50 vol % PVDF-TrFE [18]. The dielectric constant increases as the CCTO content increases in the polymer and decreases as the frequency increases [19-21]. The reason for increased low frequency dielectric dispersion was attributed due to high dielectric loss associated with CCTO [19,20]. A three phase percolative composite with Aluminum powder yielded a high dielectric constant as high as 700 [19]. The PVDF/CCTO composite processed by melt mixing and hot pressing process exhibited dielectric constant of 50 @ 100Hz with 60% CCTO content [20].

By suitably incorporating the inorganic nano particles into the host polymer, one could obtain new materials with a complimentary behavior between the two. Polymer / inorganic nano particle composites have attracted considerable attention in recent years [23-25]. Owing to their synergic properties, particularly, conducting polymer / inorganic nano crystal composites were developed for various technological applications such as capacitors, energy storage, and charge storage devices, optical and magnetic property based devices. Polyaniline (PANI) is one of the most attractive conducting polymers extensively studied due to its low cost, easy processability, environmental stability and good electrical conductivity [26-32]. A review on conducting polymer nanocomposites provided an account of the various approaches towards the synthesis and characterization of these materials [27]. Nanocomposites of polyaniline (PANI) have been widely studied by employing various techniques and the morphology of the most composites reported comprising of encapsulation of nano $TiO_2$ particles inside the shell of conducting PANI [29-35]. A dielectric constant as high as 3700 has been achieved for PANI-$TiO_2$ nanocomposite [36]. The improved conductivity of PANI-$TiO_2$ nanocomposite was also observed [37,38]. However, the PANI-$TiO_2$ composite processed by the ultrasonic irradiation reported to exhibit a decrease in conductivity with increase in $TiO_2$ content in PANI [29]. It is reported that for the PANI-$Nb_2O_5$ nanocomposite system, the conductivity, the dielectric constant and the dielectric loss decrease with increasing wt % of $Nb_2O_5$ [39]. The conductivity behaviour of the PANI-metal oxide composites have been well documented in the literature and has been found that the variation in the conductivity phenomena of the composites studied, depends on the type and content of the of fillers that are used [40-43]. The nanocomposites exhibiting both conducting and magnetic properties were reported for the PANI/$\gamma$-$Fe_2O_3$ and PANI/$Fe_3O_4$ systems [44-50]. PANI intercalated nanocomposite were reported to show improved electrical conductivity [51,52]. The dielectric properties of PANI-$BaTiO_3$ at microwave frequencies showed that even at low concentration of PANI (5 wt %), there is a drastic reduction in the dielectric constant indicating the role played by the PANI [53]. Variation in the saturation magnetization (M$s$) and coercivity (H$c$) as a function of $LiNi_{0.5}La_{0.08}Fe_{1.92}O_4$ content has been studied for the core-shell structured $LiNi_{0.5}La_{0.08}Fe_{1.92}O_4$-PANI nanocomposites [54]. There exists reports dealing with high dielectric constant composites of PANI and PANI/polymer blends [55-61]. PANI/polyurethane based composite was reported to have exhibited a dielectric constant of 1120 at 1kHz [55]. The dielectric constant ranging from 200-1000 at 1kHz was reported for the PANI/PVA composite, where the dispersed PANI particles of submicron size were suspended in the insulating polyvinyl alcohol matrix [56]. PANI incorporated in PVDF-TrFE based terpolymer [57] has shown improved dielectric constant of 1000 at 1kHz. A PANI/Epoxy blend has also been studied and exhibited a high dielectric constant close to 3000 [58].

Dielectric constant as high as $10^4$ was reported for partially crystalline PANI [59]. A high dielectric constant $>10^4$ has also been reported for the PANI system containing large microcrystalline domains (hyperbranched) [60]. Recently, high dielectric constant of $2.0 \times 10^5$ (at 1kHz) was reported for the PANI/Poly(acrylic acid) composite [61]. These findings suggest that, PANI could be effectively employed to combine ceramic in order to achieve improved electrical and dielectric properties. Incorporation of the nanofillers could enhance the electrical and dielectric properties of the polyaniline. Synthesizing materials with such large dielectric constants has been challenging and very important for the development of new generation miniatured capacitors.

To the best of our knowledge, no work pertaining to PANI/CCTO nano particle composite was reported in the literature. A simple method was used for the in-situ synthesis of PANI-CCTO nano crystal composites and characterized for their structural and dielectric properties.

## 2. Experimental

## 2.1. Materials and Characterization.

TiCl$_4$ (titanium tetrachloride, 99.98%) (Merck, Germany), calcium carbonate (BDH; A.R.grade), cupric chloride (Fluka, proanalyse grade), oxalic acid (S.D. Fine Chemicals, analytical grade), ethanol or acetone (Nice, India; 99.5% pure), Aniline monomer as received (E-merck, India), Ammonium persulphate (NH$_4$)$_2$S$_2$O$_8$ (APS), HCl (S.D.Fine chemicals) and SCMC (Sodium Carboxymethyl Cellulose) were used for the preparation of PANI-CCTO composites.

X-ray powder diffraction studies were carried out using an X'PERT-PRO Diffractometer (Philips, Netherlands) using Cu K$\alpha_1$ radiation ($\lambda$ = 0.154056 nm) in a wide range of 2$\theta$ (5$^o$ ≤ 2$\theta$ ≤ 85$^o$) with 0.0170 step size using the 'Xcelerator' check program. Infrared spectra were recorded using a Perkin-Elmer FTIR spectrophotometer employing KBr disc technique. Thermal analyses (TG) were done using the TA Instruments (UK), Model: TGA Q500, with alumina as the reference material. The experiments were carried out at a heating rate of 10$^o$C min$^{-1}$ in flowing N$_2$ atmosphere (flow rate:50cm$^3$ min$^{-1}$). Transmission electron microscopy (TEM) were carried out using FEI-Technai TEM (G-F30, Hillsboro, USA). Scanning electron microscope (SEM) (Cambridge Stereoscan S-360) was employed to study the surface morphology of the composites. The powder was cold pressed into the pellets of 12 mm diameter and 2 mm in thickness, at a pressure of 300 kg/cm$^2$. The capacitance measurements on gold sputtered pellets in two terminal parallel plate capacitor mode were carried out as a function of frequency (100Hz–1MHz) using an impedance gain-phase analyzer (HP4194A). The dielectric constants were evaluated using the standard relation $\varepsilon_r = C \times d / \varepsilon_o A$, where $C$ =capacitance, $d$ is the thickness of the pellet, $\varepsilon_o$ = 8.854X10$^{-12}$ F/m and A is the effective area of the sample.

## 2.2 Preparation of $CaCu_3Ti_4O_{12}$ nanoparticles.

$CaCu_3Ti_4O_{12}$ (CCTO) nanoparticles were synthesized using complex oxalate precursor method [62]. In a typical preparation, titania gel was prepared from the aqueous $TiOCl_2$(0.05M) by adding $NH_4OH$ (aq) (at 25°C) till the pH reached ~ 8.0 and $NH_4Cl$ was washed off on the filter funnel. This gel was added to 0.4 or 0.8 moles of oxalic acid (2 M solution) (1:1 or 1:2 ratio of Ti: $C_2O_4^{2-}$) which was kept warm (~40°C). To the clear solution obtained, calcium carbonate was added in aliquots and stirred. An aqueous solution containing titanyl oxalic acid together with calcium titanyl oxalate remained clear without any precipitate formation. This solution was cooled to 10°C to which cupric chloride dissolved in acetone along with water (80/20) was added and stirred continuously. The thick precipitate was separated out by further addition of acetone. Subsequently, the precipitate was filtered, washed several times with acetone to make it chloride-free and dried in air. The precursor was isothermally heated around 700°C to get nanoparticles (20-75nm) of phase-pure calcium copper titanate, $CaCu_3Ti_4O_{12}$ as confirmed by X-ray diffraction and TEM studies.

## 2.3 Preparation of PANI-$CaCu_3Ti_4O_{12}$ nanocomposites (In-situ polymerization)

The most preferred method for the synthesis of PANI is to use either HCl or $H_2SO_4$ with ammonium peroxydisulfate as an oxidant [63,64]. The PANI-$CaCu_3Ti_4O_{12}$ nano crystal composites were prepared by in-situ polymerization in dil.HCl. For the preparation of PANI/CCTO nano crystal composites, as a first step, CCTO nanoparticles were added to sodium carboxyl methyl cellulose (SCMC), which acts as dispersant, in the ratio 1.5 (SCMC):100(CCTO) in 100ml dil. HCl (pH 1-2) and ultrasonicated. The ultrasonication is to avoid the agglomeration and to set the nanoparticles wet with the SCMC. To the above suspension, 2 gm of aniline monomer was added and stirred for 1-2 h. 7.35gm of APS in 30 ml of 1M HCl solution was added drop wise to the above mixer and stirred for 12 h at room temperature using silicone oil bath. The dark green solid mass thus prepared was washed several times with dil. HCl. Then it was further washed with ethanol and filtered, dried in vacuum oven at 80-100°C for 24 h. Composites with varying concentrations of CCTO (10, 30 and 50 %) by weight were fabricated.

The polyaniline (PANI) that is characterized in the present investigation has been prepared by similar method that is described above.

**3.0 Results and discussion**

*3.1. X-ray diffraction studies*

The X-ray powder diffraction patterns obtained for pure PANI, pure CCTO and series of PANI-CCTO composites with different weight fractions of CCTO are shown in Fig.1. In Fig.1(a),the diffraction peak at $2\theta = 25°$ (200) is typical of PANI [65] apart from the peaks corresponding to $2\theta = 43.8, 49$ and $50.9°$ which are attributed to the nanocrystalline nature of PANI [66]. Fig.1(b) shows the X-ray diffraction pattern obtained for the pure CCTO nano crystalline powder which is compared well with the ICDD data (01-075-1149) and with that reported in our previous publication [62]. The X-ray diffraction pattern obtained for the various composites are depicted in the Fig.1(c-e). It is evident from these figures that as the CCTO content increases in the PANI, the peaks corresponding to CCTO dominate. The peak at $25°$ (200) corresponds to PANI (Fig. 1c) for the composite (90% PANI+10%CCTO) is less intense as compared to that of the pure PANI (Fig.1a). The intensity (Fig.1d) has further diminished for the composite corresponding to 70% PANI+30%CCTO.

*3.2. FTIR analysis*

The FTIR spectra recorded for pure PANI, nano crystalline CCTO and their composites are given in Fig.2. As shown in Fig.2(d), the characteristic peaks of PANI occur around 1567,1484,130,1242 and 1118 cm$^{-1}$. The absorption peaks at 1567 and 1484 cm$^{-1}$ are attributed to the C = C stretching of the quinoid and benzenoid rings [67,68]. The peaks at 1301 and 1242 cm$^{-1}$ are assigned to the C-N stretching modes of the benzenoid rings [68], while the band at 1118 cm$^{-1}$ is assigned to an in plane bending vibration of C-H (mode of N=Q=N, Q=N$^+$H-B and B-N$^+$H-B) which formed during the protonation [69,70]. The peak at 801 cm$^{-1}$ is attributed to the out of plane deformation of C-H in the p-disubstituted benzene ring [71]. The IR spectrum was recorded for the CCTO nano particles and is depicted in Fig.2(e). The absorption bands in the region 380–700 cm$^{-1}$ arise from the mixed vibrations of CuO$_4$ and TiO$_6$ groups prevailing in the CCTO structure. The IR spectra recorded for the composites (Fig.2 (a-c)), exhibit the spectra similar to that of pure PANI unlike in the case of X-ray diffraction wherein the characteristics of both PANI and CCTO are reflected.

### 3.3 Thermogravimetric analysis

Thermogravimetric analyses (TGA) of pure PANI and the PANI-CCTO nanocrystal composites were done to assess their thermal stability and the results obtained are shown in Fig.3. As reported [72], PANI has exhibited three typical steps of weight loss behaviour. In the first step, a small fraction of weight loss occurred around 110$^o$C which is attributed to the loss of water molecule absorbed in the sample [72-74]. The second step around 285$^o$C is for the coevolution of water, acid or phase transition [72]. The third step could not be clearly deciphered, however, degradation of PANI takes place after 285$^o$C. The observed weight loss for PANI at 590$^o$C is around 39% and the weight loss decreases as the weight fraction of the CCTO increases in the PANI-CCTO composite. This behaviour is similar to that reported in the literature for PANI based composites [39]. The TGA in essence indicate that the composites have better thermal stability than that of pure PANI.

### 3.4 SEM/TEM Studies

The SEM micrographs obtained for PANI are shown in Fig.4. The micrograph (left) showing the agglomerated PANI, which is on careful examination at high magnification (right) reveals the presence of the existence of nanofibers in the polymer. Fig. 5(a-c) shows the bright field TEM images of nano powders of CCTO and SAED pattern of CCTO. Fig. 5(a,b) presents the bright field TEM images of the CCTO nano powders obtained from the oxalate precursor and the size of the crystallites is in the range of 20-75 nm. Fig.5(c) shows the SAED pattern with the zone axis as [012]. SAED pattern confirms the crystalline nature of the particles. The ratio of the reciprocal vectors ($t_2/t_1$) is around 1.229, approaching the calculated value of 1.225 for the bcc lattice. Fig. 6(a) shows the bright field TEM image of PANI exhibiting nanofibrous morphology and observed that the dimensions of the PANI nanofibre are ranging from 150-200 nm in width and the aspect ratio could not be specified as there is some agglomeration and non-uniform thickness associated with the fibre. The dimensions of the PANI and CCTO were determined and observed that (Fig. 6(b)), the width of the PANI fibre is around 300 nm (marked in 6(b)) and the size of the CCTO crystallites embedded in the PANI are in the range of 30-50 nm. The Fig.5(a-b) obtained for CCTO particles is of high magnification wherein which one could see fusing of nanocrystallites. Figs.6(a) and 6(b) are obtained for pure PANI and the PANI-CCTO composite respectively. The magnification is relatively lower than that involved in Fig.5(a-b). Fig.6(b) illustrates only the magnified version of PANI fibres embedded with CCTO crystallites. Since Fig.6(b) is obtained at higher magnification, one could take a close look at the fibre. As reported, PANI nanofibers are shown to form without any template or surfactants. In this work we also found that some population of nanofibers occurs spontaneously during the chemical oxidative polymerization of aniline without the addition of any structural directing agents as reported by Huang and Kaner [75].

### 3.5. Dielectric properties and AC conductivity

The log-log plot of frequency dependent dielectric constant ($\varepsilon_r$) at room temperature is shown in Fig.7. The inset shows the dielectric constant as a function of weight percent CCTO at different frequencies. As shown in Fig.7, high dielectric constant ($>10^6$) has been obtained for the PANI synthesized in this work. The dielectric constant obtained for PANI in this work is much higher than that reported in the literature [59,60]. It maybe owing to the presence of nano crystalline domains rather than micro domains present in the hyper branched PANI. The observed high dielectric response ($>10^6$) for the nano crystalline PANI at low frequencies (100Hz-1kHz) may be correlated to the hyper-electronic polarization [76] and a strong polaron delocalization [59,77].

All the composites exhibited higher dielectric constants ($\varepsilon_r > 10^6$) than that of pure PANI and CCTO. The dielectric constant increases as the CCTO content increases in the PANI. It is also reported that the introduction of nanoparticles into the PANI matrix, increases the dielectric constant of the composite [36,44]. However, the dielectric constant decreases with the increases in frequency, which is a known trend akin to that reported in the other PANI/composites [55,56]. The $\varepsilon_r$ value obtained for all the three PANI/CCTO composite samples lies in the range of 3.1-4.6x$10^6$ at 100Hz. At 100kHz, the $\varepsilon_r$ value obtained is $> 10^5$. It is to be noted that the $\varepsilon_r$ value obtained for 50% PANI+ 50% CCTO composite is higher than that of 90% PANI+10% CCTO at all the frequencies under study. The $\varepsilon_r$ value obtained for the PANI-CCTO composite in this work is much higher than that reported for the other PANI based composite systems [55-58,61]. The CCTO nanocrystallites surrounded by the network of PANI acting as nanodielectric domains yielded high dielectric constants via the interfacial /Maxwell Wagner type of polarization mechanism.

The dielectric loss recorded as a function of frequency is shown in Fig.8. The inset shows the dielectric loss as a function wt. % of CCTO at different frequencies. The dielectric loss for PANI has very high value of 18.7 at 100Hz, which decreases drastically to 1.2 around 100kHz and subsequently, increases to 2.2 at 1MHz. The composites also exhibit a similar trend to that of PANI. The observed loss value for 90% PANI+10% CCTO is 2 times less than the value obtained for pure PANI around 100Hz. Similar trend has been observed for the other composites also. However, the dielectric loss decreases gradually as the frequency increases upto 1kHz and sharply increases at higher frequencies (Fig.8).

The dielectric behaviour was rationalized using the universal dielectric law (UDR), according to this model [78], $\varepsilon'$ could be written as

$$\varepsilon' = \tan(s\pi/2)\sigma_o f^{s-1}/\varepsilon_o \qquad \text{---1}$$

where $\sigma_o$ and frequency component $s$ are temperature dependent, $\varepsilon_o$ is the permittivity of free space. Equation (1) can be rewritten as

$$f\varepsilon' = A(T)f^s, \qquad \text{---2}$$

with temperature-dependent constant $A(T) = \tan(s\pi/2)\sigma_o/\varepsilon_o$.

The above equation in logarithmic form could be written as

$$\log(f\varepsilon') = \log(A(T)) + s\log f \qquad \text{---3}$$

Therefore, for a given temperature, a straight line with slope of $s$ could be obtained if $\log_{10}(f\varepsilon_r')$ is plotted against $\log_{10}(f)$, which is evidenced in Fig.9. The value for the parameter $s$ deduced from the linear fit for the CCTO is 0.95, which is in agreement with that reported in the literature [79] while that ($s$) for the PANI is 0.59 and for PANI-CCTO nanocrystal composite is 0.63. The exponent $s$ is connected with the ordering within the cluster and small values of $n$ correspond to the highly irregular clusters whereas large values of $n$ correspond to a highly ordered structure [80]. Our data reveal that $n$ is low for pure PANI and increases with the CCTO content in the composites, which shows that the addition of CCTO to PANI increases the order. It is envisaged that polymer chains organize themselves into more ordered structures with the addition of CCTO and hence the overall increase in the order. The dielectric behaviour of CCTO, PANI and the composites follow the universal dielectric law (UDR).

The frequency dependent AC conductivity behaviour of PANI/CCTO composites is shown in Fig.10. The inset in Fig.10 indicates the frequency dependent AC conductivity of PANI and CCTO. The AC conductivity of PANI is in the order of $5 \times 10^{-5}$ s/cm which is attributed to the low level of protonation as reported [39], and the composites exhibited higher AC conductivity response than that of the pure PANI. The conductivity for PANI (inset in fig.10) increases gradually upto 100kHz and increases sharply beyond 100kHz, which is consistent with the observed high frequency dielectric loss response. All the composites exhibited a similar trend, but the conductivity is almost constant upto 2kHz and then increases as the frequency increases. In this work, it is observed that the conductivity increases slightly upto 2kHz as the CCTO content increases in the PANI which is in contrast with that observed for PANI-$Nb_2O_5$ composite [39]. This type of trend has been observed in the other PANI based systems [37,38,40] and is attributed to the polarization of the charge carriers and may also be due to the existence of interface nano-trapping levels effectively interacting with the polaronic states. At higher frequencies, the conductivity decreased as the CCTO content increased, which is the characteristic feature of the disordered materials and due to the contribution of polarons, which move smaller distances in the amorphous region and this supports the presence of isolated polarons in this region [42,43].

## 4. Conclusions

High dielectric constant Polyaniline (PANI) / $CaCu_3Ti_4O_{12}$(CCTO) nanocrystal composite was synthesized using in-situ polymerization of aniline in dil.HCl. The composite corresponding to 50%CCTO-50%PANI exhibited higher dielectric constant ($4.6 \times 10^6$ @100Hz). As revealed by the TEM, CCTO nanocrystals are embedded in the nanofibers of polyaniline matrix , which in turn act as nanodielectric domains in the composites which could be the source for the high dielectric response of the composite. PANI has been considered as one of the most potential conducting polymers due to its high conductivity. On the other hand, CCTO has gained considerable attention due to its unusual high dielectric constant ($\varepsilon \sim 10^{4-5}$) which is nearly independent of frequency (upto 10 MHz) and low thermal coefficient of permittivity (TCK) over a wider range of temperature (100-600K). In this work, it is observed that, as the CCTO content increases in PANI, the dielectric constant increases but the loss decreases implying that one could tune the dielectric properties by adding the known quantities of CCTO to PANI. This would facilitate the design and fabrication of PANI based materials for the potential applications in light emitting diodes, information storage, frequency converters, modulators, dielectric amplifiers, sensors, anti corrosion coatings, lightweight battery electrodes and electromangnetic shielding devices. These composite with suitable binders, could also be used as electrode materials for the super capacitor applications as demonstrated in the case of the other oxides [81,82]. Investigations have been in progress for the development of electrode materials using the PANI/CCTO composites using PVDF as binder.

**Acknowledgements**

The management of Central Power Research Institute is acknowledged for the financial support (CPRI Project No.5.4.49).

**Figure captions:**

**Figure. 1.** The XRD diffraction pattern for : (a) PANI, (b) CCTO, (c) ) 90% PANI+10% CCTO (d) 70% PANI+ 30% CCTO, and (e) 50% PANI + 50% CCTO.

**Figure. 2.** FTIR spectra for : (a) 50% PANI + 50% CCTO, (b) 70% PANI + 30% CCTO (c) 90% PANI + 10% CCTO, (d) PANI, and (e) CCTO.

**Figure. 3.** TGA curves for: (a) PANI, (b) 90% PANI+10% CCTO (c) 70% PANI+ 30% CCTO, and (d) 50% PANI/ 50% CCTO.

**Figure.4.** SEM micrographs showing the powders (left) are agglomeration of nanofibers (right).

**Figure.5.** (a,b). Bright field TEM images of phase pure CCTO crystallites dimensions ranging from 20-75 nm and (c) SAED pattern obtained with the zone axis of [012], $t_2/t_1=1.229$.

**Figure. 6.** (a) Nanofibrils nature of PANI, dimensions ranging from 150-200 nm and (b) 50% PANI-50% CCTO nano composite showing CCTO nanocrystallites (30-50nm) embedded in the PANI matrix (300nm).

**Figure.7.** Frequency dependent dielectric constant ( log-log scale) of PANI and PANI/CCTO nano crystal composites.

**Figure 8.** Frequency dependent dielectric loss for PANI, CCTO and PANI-CCTO nano composites.

**Figure.9.** Plot of $\log_{10}(f\varepsilon_r^{'})$ against $\log_{10}(f)$ for the CCTO, PANI and 50% PANI+50%CCTO nano composite at room temperature.

**Figure.10.** Frequency dependent AC conductivity PANI-CCTO composites and the inset for pristine PANI and pure CCTO.

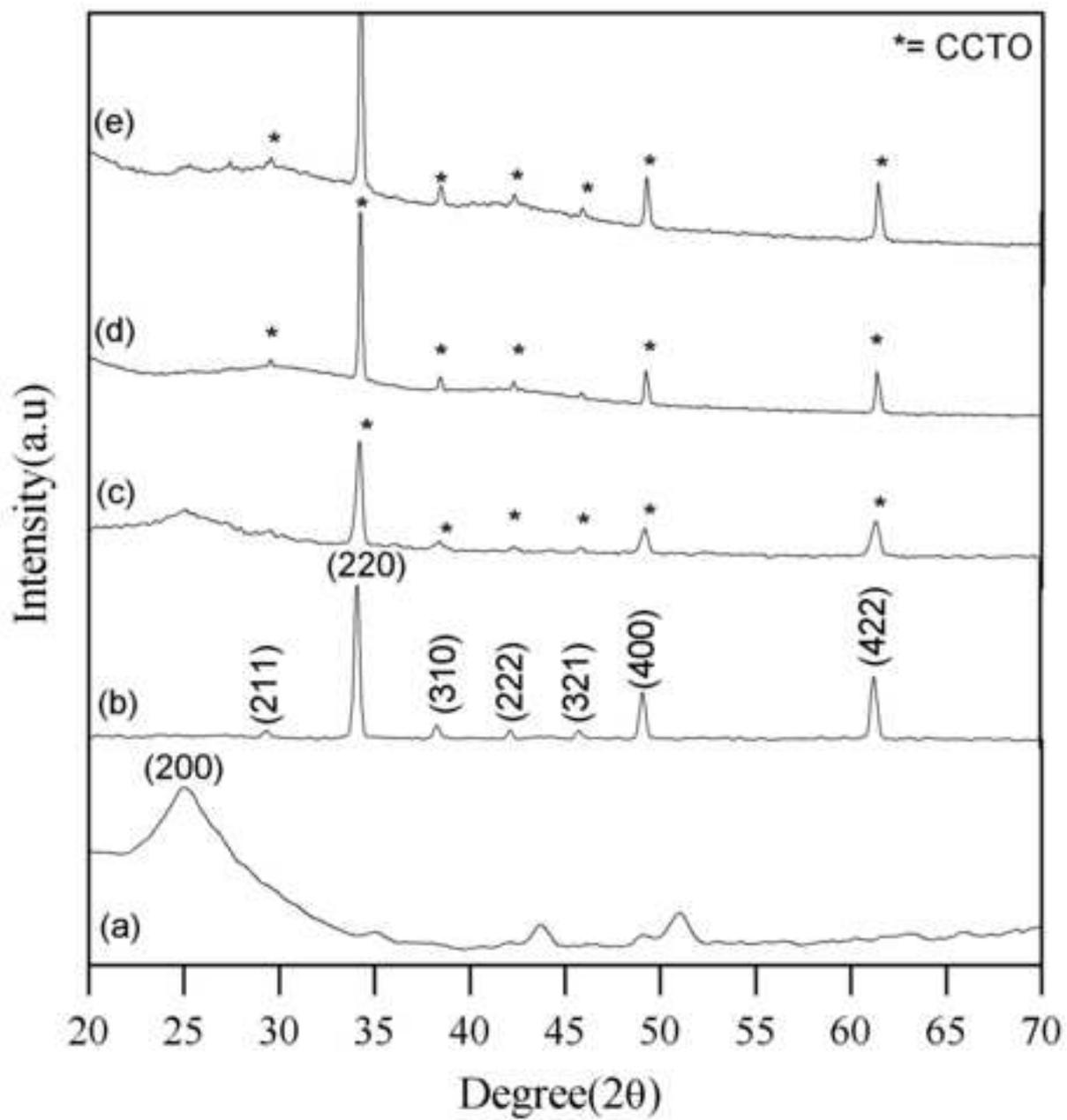

**Figure 2**
**Click here to download high resolution image**

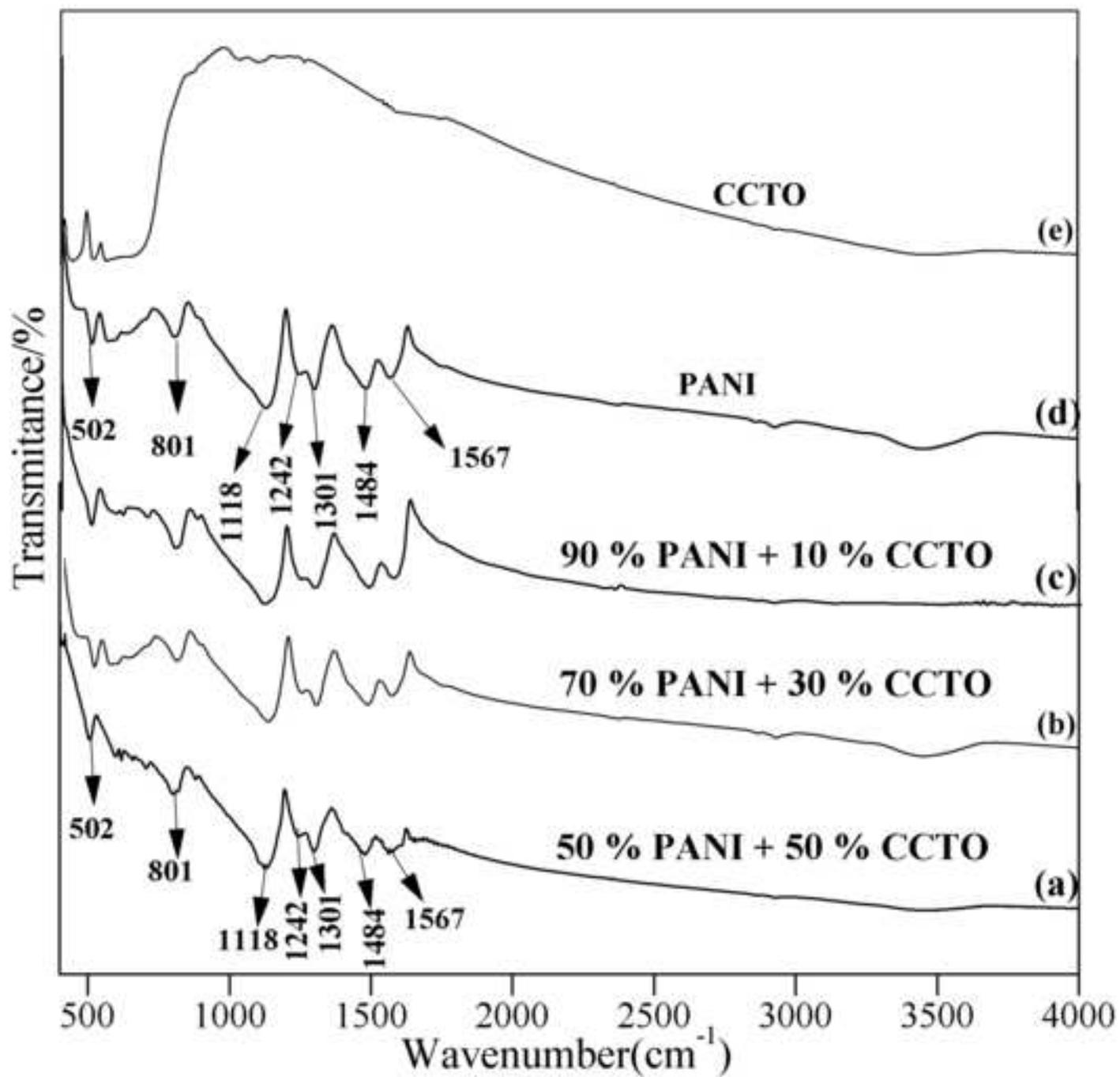

**Figure 3**
**Click here to download high resolution image**

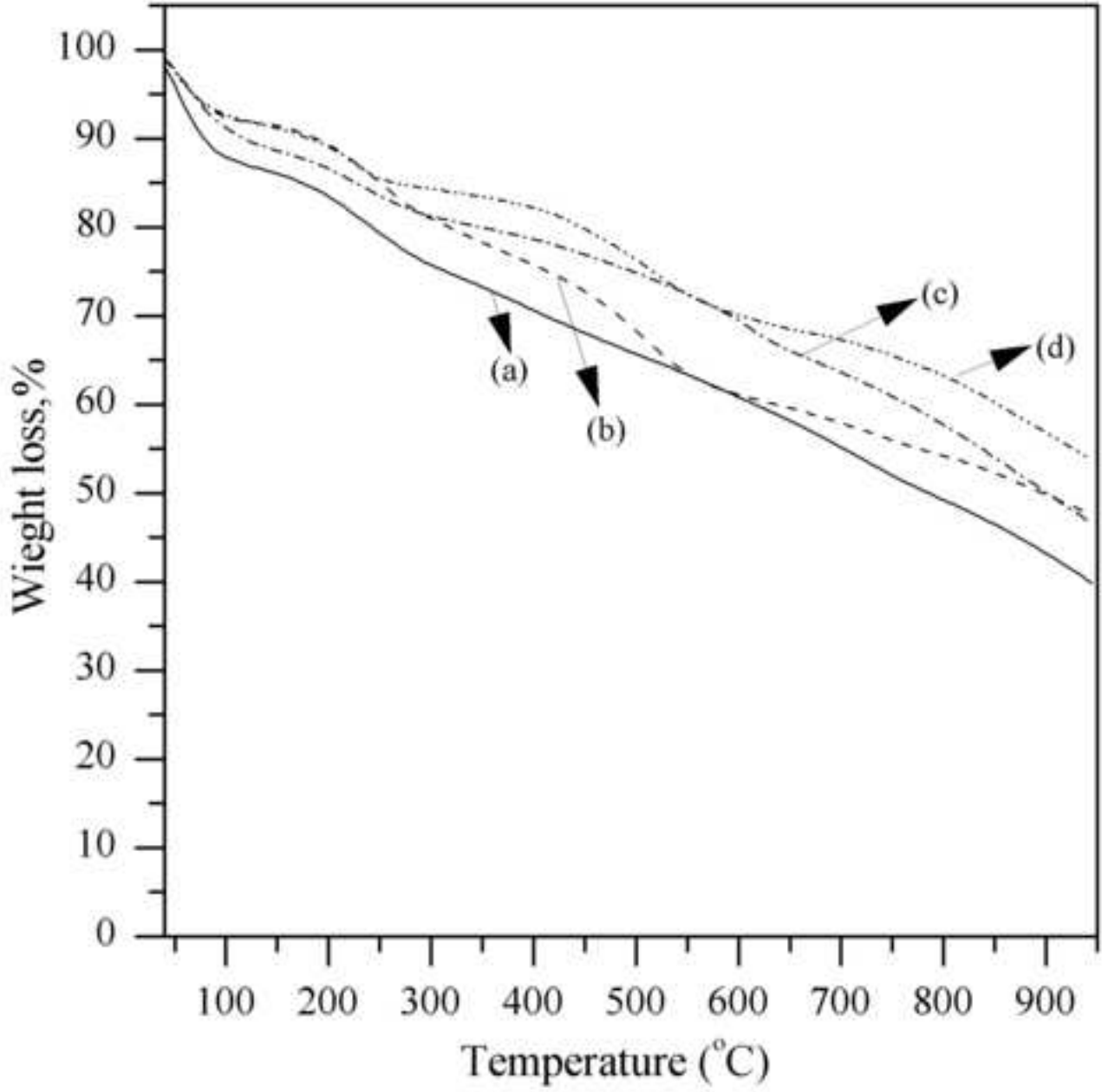

**Figure 4**
Click here to download high resolution image

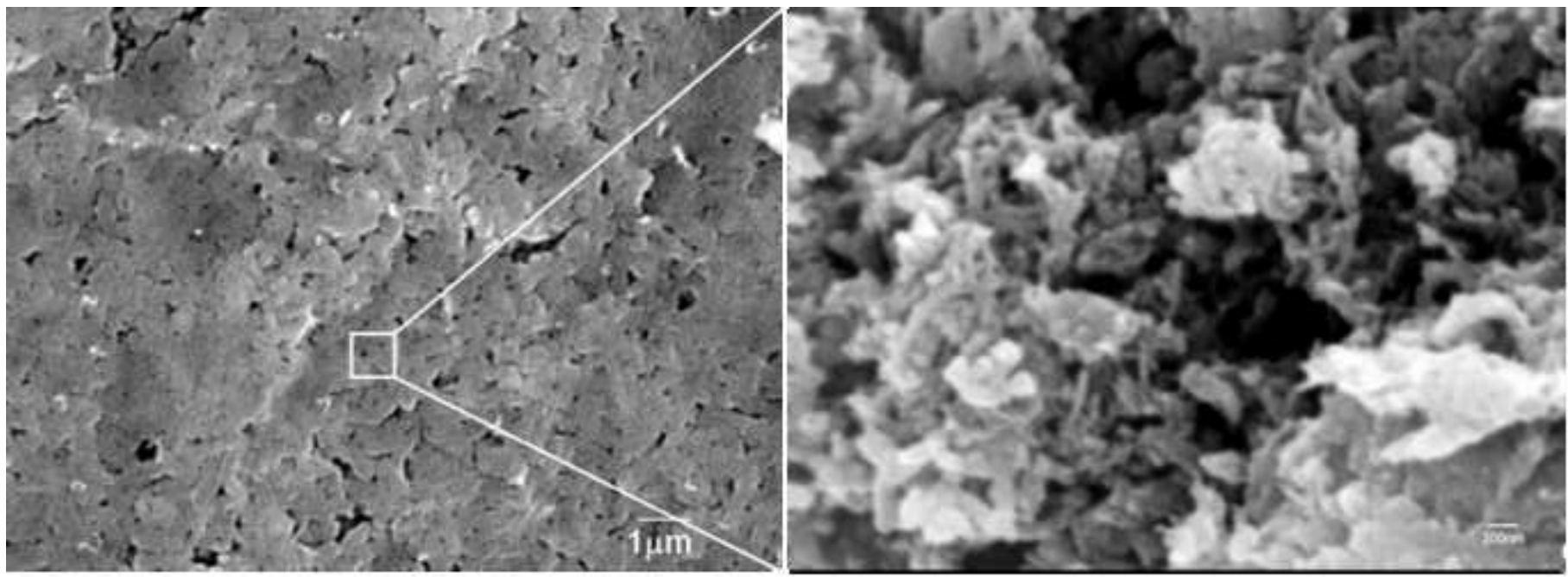

**Figure 5**
**Click here to download high resolution image**

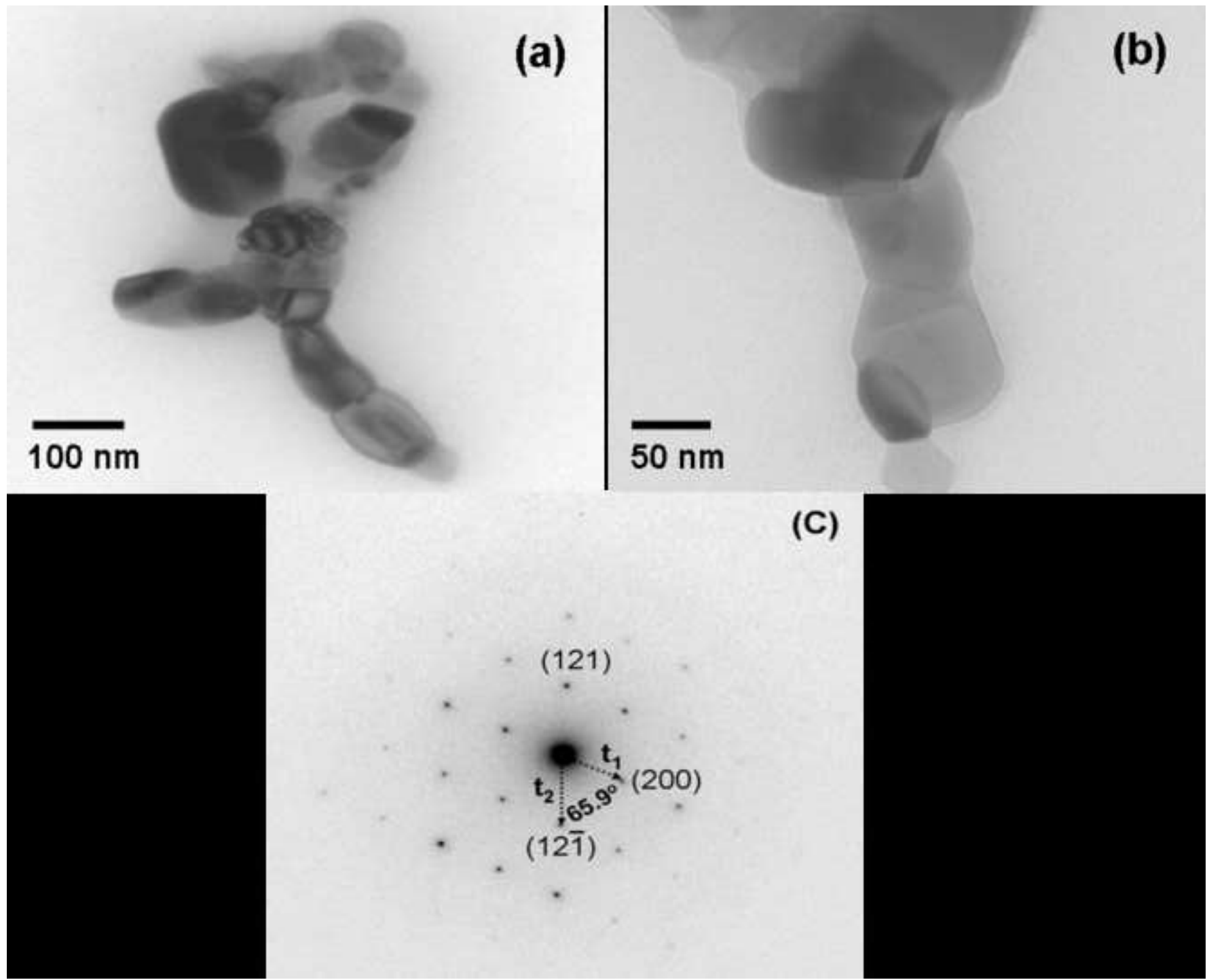

**Figure 6**
[Click here to download high resolution image]

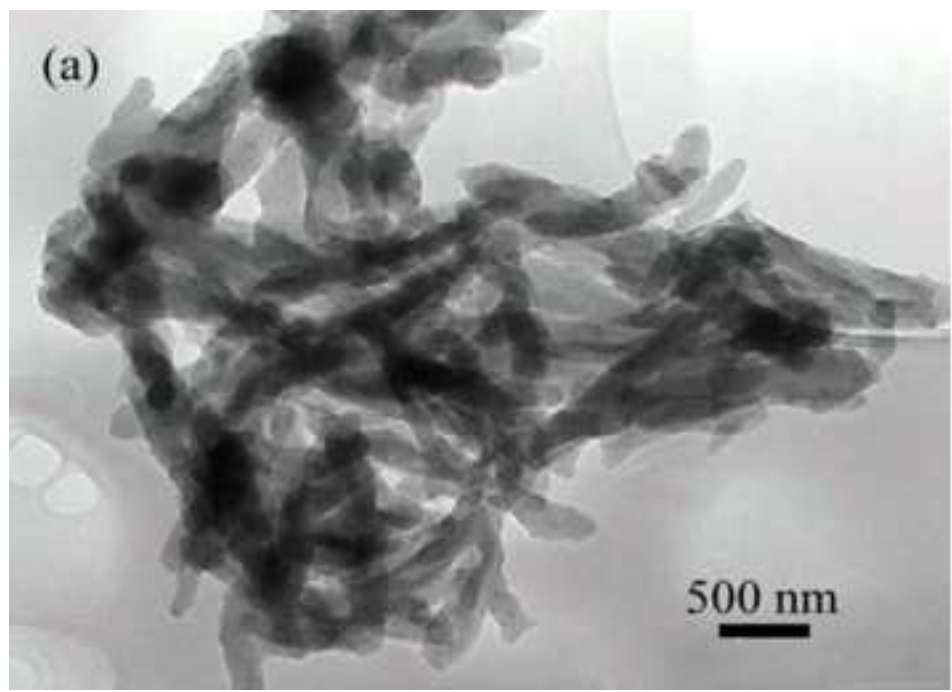 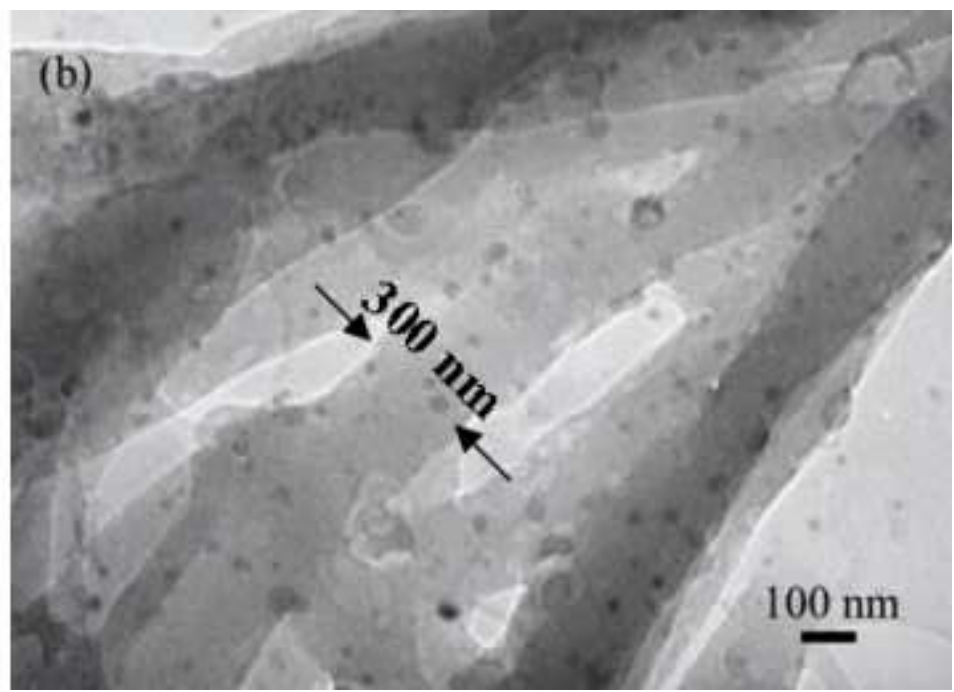

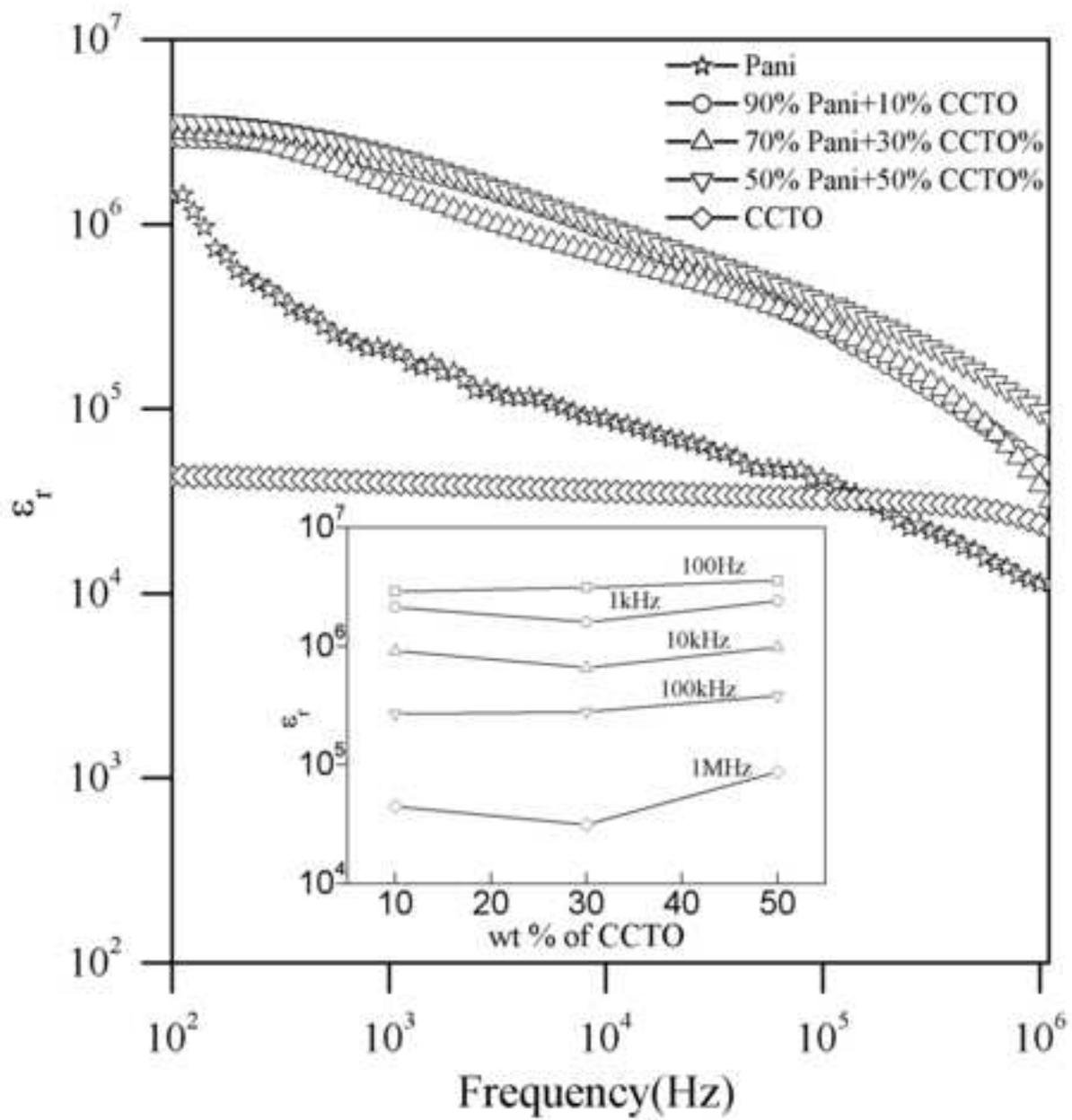

**Figure 8**
Click here to download high resolution image

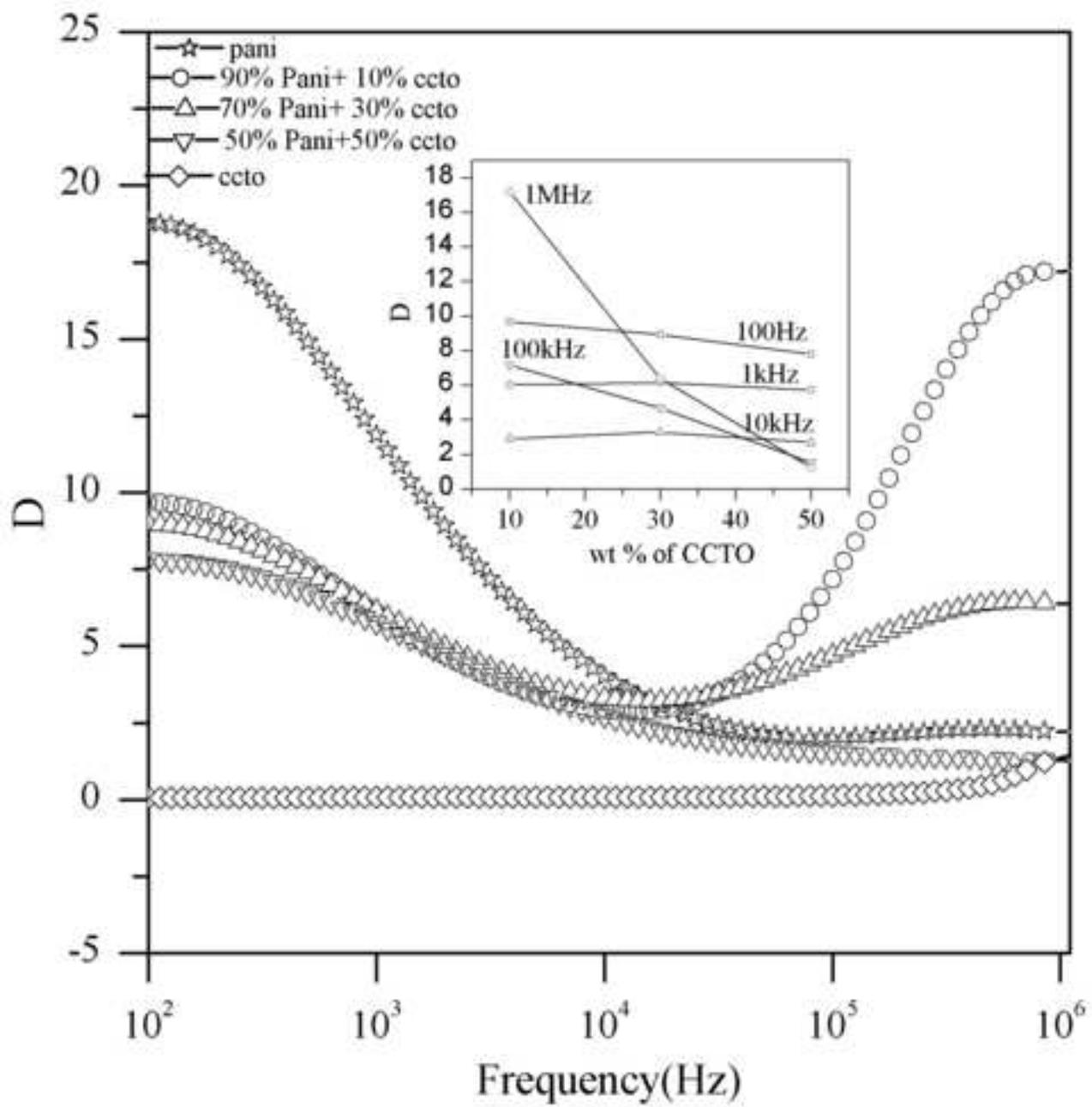

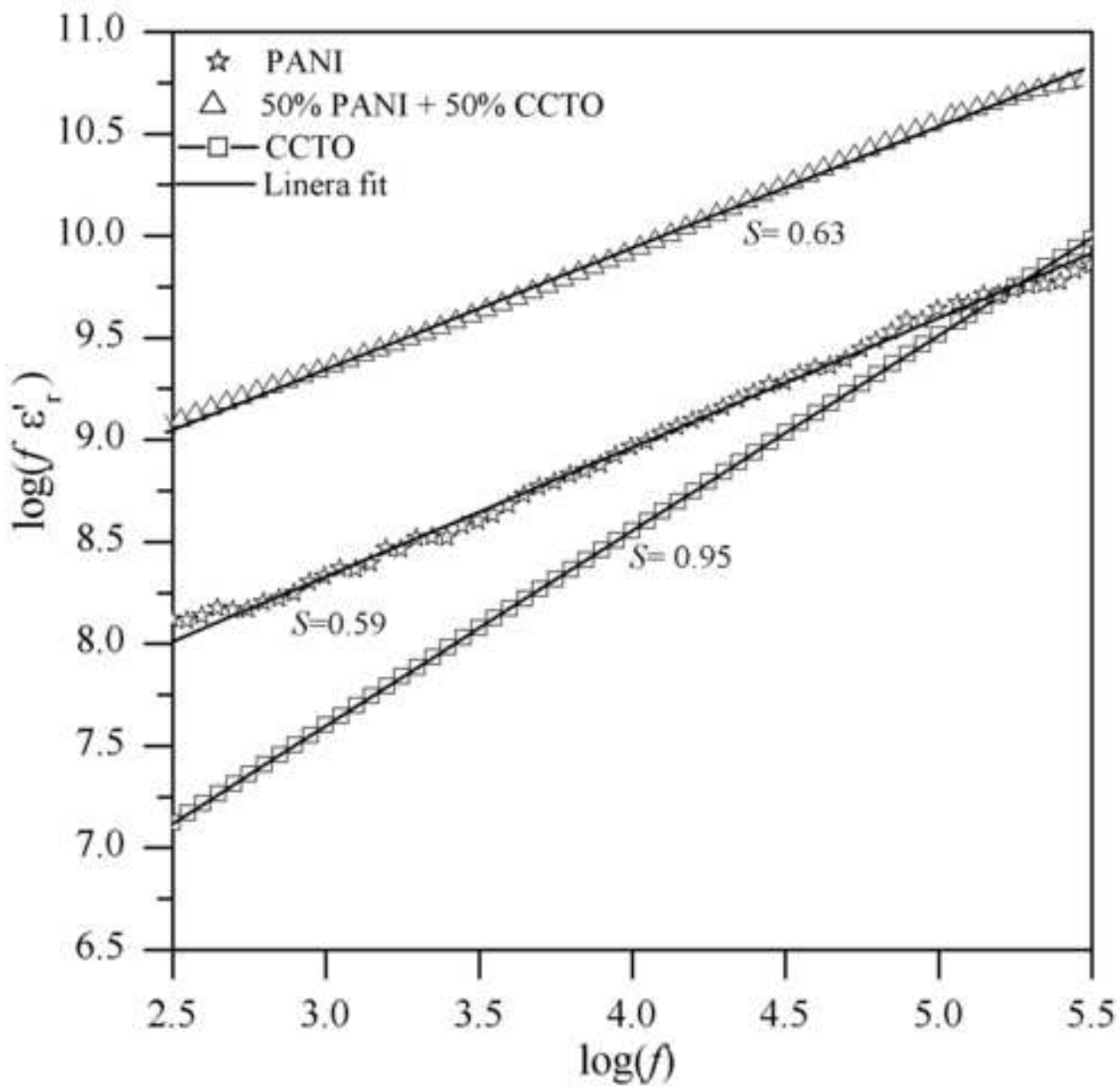

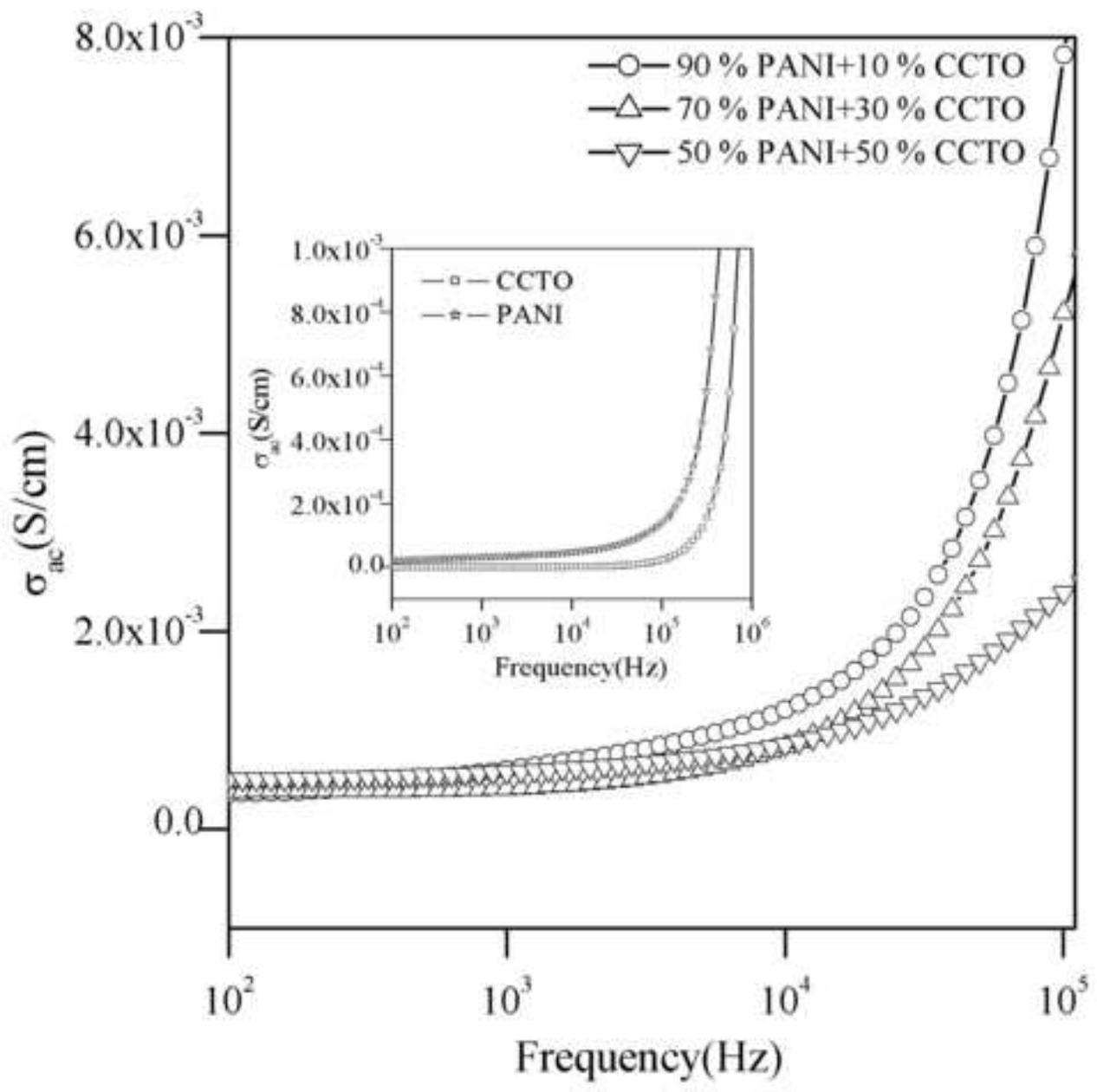